\documentstyle[12pt,epsf]{article}

\newcommand{\r}{\rightarrow}

\newcommand{\be}{\begin{equation}}
\newcommand{\ee}{\end{equation}}
\newcommand{\eel}[1]{\label{#1}\end{equation}}
\newcommand{\bea}{\begin{eqnarray}}
\newcommand{\eea}{\end{eqnarray}}
\newcommand{\eeal}[1]{\label{#1}\end{eqnarray}}
\newcommand{\baq}{\begin{equation}\begin{array}{rcl}}
\newcommand{\eaq}{\end{array}\end{equation}}
\newcommand{\eaql}[1]{\end{array}\label{#1}\end{equation}}
\newcommand{\beac}{\begin{equation}\begin{array}{rcl}}
\newcommand{\eeacn}[1]{\end{array}\label{#1}\end{equation}}
\newcommand{\ba}{\begin{array}}
\newcommand{\ea}{\end{array}}
\newcommand{\non}{\nonumber \\}

\renewcommand{\a}{\alpha}

\newcommand{\al}{{\alpha^{'}}}
\newcommand{\beq}{\begin{eqnarray}}
\newcommand{\eeq}{\end{eqnarray}}
\newcommand{\nn}{\nonumber}

\newcommand{\w}{Schwarzschild $\:$}

\newcommand{\bl}{\hspace{-.65cm}}
\newcommand{\gym}{g_{YM}}
\newcommand{\gef}{g_{eff}}
\begin{document}
\newcommand{\preprint}[1]{\begin{table}[t]  
           \begin{flushright}               
           \begin{large}{#1}\end{large}     
           \end{flushright}                 
           \end{table}}                     
\preprint{hep-th/9802042\\TAUP-2474-98 \\ HUTP-98/A003}

\begin{center}
\LARGE{Supergravity and The Large $N$ Limit of Theories \\
With Sixteen Supercharges}

\vspace{10mm}

\normalsize{Nissan Itzhaki${}^{1}$,  Juan M. Maldacena${}^{2}$, \\ 
Jacob Sonnenschein${}^{1}$ and Shimon Yankielowicz${}^{1}$}

\vspace{10mm}

{\em ${}^{1}$ Tel Aviv University, Ramat Aviv, 69978, Israel\\
sanny, cobi, shimonya@post.tau.ac.il\\
${}^{2}$  Lyman Lab., Harvard University, Cambridge, MA 02138,
USA\\
malda@pauli.harvard.edu}
\end{center}

\vspace{10mm}

\begin{abstract}
We consider field theories with sixteen supersymmetries, which 
includes U(N) Yang-Mills theories in various dimensions, and 
 argue that their large $N$ limit is related to certain supergravity
solutions. 
We study this by considering a system of D-branes in string theory 
and then taking a limit where the brane worldvolume theory decouples from
gravity. At the same time we study the corresponding D-brane
supergravity solution and  argue that we can trust it in certain 
regions where the curvature (and the effective string coupling,
 where appropriate) are 
small. The supergravity solutions typically have
several weakly coupled 
regions and interpolate between different limits of string-M-theory.

\end{abstract}

\newpage

\baselineskip 18pt

\section{Introduction}

String theory contains D-branes which are solitonic objects \cite{polchinski}. 
When we consider the full theory in the presence of these solitons
we have modes that propagate in the bulk and modes that propagate
on the solitons. The modes on the soliton interact with each other
and with the bulk modes. 
It is possible, however, to define a limit of the full theory in which
the bulk modes decouple from the modes living on the D-brane. 
This is typically a low energy limit, in which we tune the 
coupling constant so as to 
keep only the interactions  among the
modes living on the D-brane. 
In this limit the D-brane theory 
becomes super-Yang-Mills (for $p\leq 3$).
Separating the branes   by some distance corresponds in the field theory
to 
giving  Higgs expectation values to some fields. 
Since we want to keep these expectation values finite when we take
the limit,  we should consider the branes at 
substringy distances \cite{kpdkps}.

Since D-branes carry some mass and charge they excite the bulk gravity modes
and we can find supergravity solutions carrying the same mass and charges.
Naively the supergravity solution describes only the long range fields
of the D-branes, since we do not expect supergravity to be valid at
short distances. General covariance, however, tells us that
we can trust the supergravity solution as long as curvatures
are locally small compared to the string scale (or the Planck scale). 
A more careful analysis shows that for a system with a 
large number of branes, large $N$, 
the curvatures are small and 
we can trust the supergravity solutions even at 
the substringy distances involved in the decoupling limit described above. 
The situation is similar to the one studied in \cite{mal} for conformal
field theories (see also \cite{dps,klebanov}).
In particular for the 4D $ N=4$ $U(N)$ super-Yang-Mills theory associated with
 $N$ D3-branes, it has been argued in \cite{mal} that it is ``dual'' to
 type IIB string theory on $AdS_5\times S^5$ in the large $N$ limit.

The aim of this paper is to explore  analogous   connections in the
more  general case of non-conformal field theories.
The supergravity solutions corresponding to $p+1$ super-Yang-Mills
are black $p$-brane solutions. They are extended along $p+1$ spacetime
dimensions. We interpret the radial variable as being related to the
energy scale of the process involved. One of the reasons for this 
interpretation is the fact that a Dp-brane sitting at some position
$r$ corresponds to giving a Higgs expectation value to some 
fields which break the gauge group $U(N+1) \rightarrow U(N)\times U(1)$. 
Large values of $r$ correspond to very large Higgs expectation values, 
which have dimension of energy,
 and therefore correspond to large energy
scales. 
Hence, large values of $r$ correspond in the field theory to the UV
region and
small values to the IR region. 
The curvature and the value of the dilaton depend on the radial variable
$r$.
The radial dependence of the dilaton represents the running of the 
effective coupling constant (which is these theories is simply given by
dimensional analysis). 
For $p<3$ the solutions have large curvatures for large values of $r$,
and we cannot trust them in this region. This is not a 
 problem since 
in the UV we can trust the perturbative description and we do not
expect perturbation theory and gravity to be valid at the same time. 
As we move to smaller values of $r$, i.e. lower energies, we find
a region with small curvature and small string coupling. 
In this region we can trust a string theory description on the 
corresponding background. All Yang-Mills theories considered here
contain strings in this sense (as in the $p=3$ case \cite{mal}). 
These examples realize the general description of large $N$ gauge theories
proposed by Polyakov \cite{polyakov}. In the language of \cite{polyakov}
 the radial variable is related to the    Liouville field. 
This string theory description is valid for intermediate values of $r$. 
In some cases we can use a dual description for the small $r$ region.
In general we have a reliable supergravity description in the large
$N$ limit only for certain energy scales (which is to be expected since
the coupling depends on the energy).

For $p>3$ we have a similar situation with the coupling running 
in the reverse direction, small r corresponds to weakly coupled 
super-Yang-Mills and for large $r$ we will have to use dual descriptions.

Since the conformal case $p=3$ is the borderline 
 case we shall discuss first
 the $p<3$ theories and then the $p>3$ theories.
The paper is organized as follows.
In Sec.2 we describe some general properties which are common to all
 Dp-branes.
In Sec.3 we consider  D2-branes and obtain a relation with the 2+1 conformal
field theory at low energies. In particular, the flow
of the super-Yang-Mills in 2+1 dimensions to a superconformal field theory 
with SO(8) R-symmetry is realized by 
the supergravity  solution of the D2 branes. 
We also relate the super-Yang-Mills with small temperature to the near 
extremal
 M2-branes solutions.
In Sec.4 we discuss D1-branes.
 We show that for large $N$ there is an intermediate region between 
perturbative super-Yang-Mills in the UV and  
the orbifold $(R^8)^N/S_N$ CFT with the Dijgraff-Verlinde-Verlinde
vertex operators \cite{dvv} in the IR.
This  intermediate region is described by type IIB on a non-trivial 
background.
In Sec.5 we study  D0 branes and the relation to black holes in 
matrix theory \cite{bfss,susskind,bhmatrix,homa}.
In Sec.6 we  consider D4 branes and the relation with the six-dimensional
(0,2) field  theory compactified on a circle. 
The flow 
of the (0,2) theory on a circle to the 4+1 dimensional
 super-Yang-Mills in the 
IR has a counterpart in the supergravity solution. 
In Sec.7  we briefly discuss the case of D5 branes and their relation to 
IIB NS 5branes and we make, in Sec.8, a digression into IIA NS fivebranes. 
Finally, in Sec.9 we consider D6 branes where we conclude 
(as in \cite{ssns,sen,sei2})
that the theory does not decouple from the bulk.
We show that a finite temperature configuration 
is described by a
\w black hole in five dimensions.  
To make the outcome of the analysis of each case  clearer, we give a short
 summary of the conclusions at the end of each section.

The connection between Yang-Mills theories and supergravity solutions
was explored following a different method in \cite{vijayetal}.

\section{Generalities}

We study D$p$-branes in the field theory limit\footnote{
We use conventions in which $g_s \rightarrow 1/g_s $ under S-duality
for type IIB. The IIA conventions are such that the radius of the
eleventh circle is $R_{11} = g_s \sqrt{\alpha'}$. 
The eleven dimensional Planck length is defined as  $l_p = 
g^{1/3} \sqrt{\alpha'}$. 
}\cite{malpro,sen,sei2}
\be\label{limit}
 \gym^2=(2\pi)^{p-2} g_s\al^{(p-3)/2}=\mbox{fixed},\;\;~~~~~
\al\r 0,
\ee
where $g_s=e^{\phi_{\infty}}$, and $\gym $ is the Yang-Mills coupling
constant. We keep the energies fixed when we take the limit. 
For $p\leq3$ this limit implies that the theory 
decouples from the bulk since the ten dimensional Newton 
constant goes to zero. It also suppresses  higher order corrections
 in $\alpha'$ to the action.
For $p>3$ we have $g_s\r \infty$ which implies that we should use a
dual description to analyze the decoupling issue.

When we take the limit (\ref{limit}) we are interested in
finite energy configurations in the field theory. 
This corresponds
to finite Higgs expectation values. 
We are, therefore, considering  the limit 
\be\label{limit2}
U\equiv \frac{r}{\al}=\mbox{fixed},\;\;\;~~~~\al\r 0.
\ee
In terms of the field theory $U$ is the expectation value of the Higgs.
Note that in this limit $\frac{r}{l_s}\r 0$ which means that we study the 
system at substringy distances. 
At a given energy scale, $U$, the effective  dimensionless coupling constant
in the corresponding super-Yang-Mills theory 
 is $g^2_{eff}\approx \gym^2NU^{p-3}$.
Thus, perturbative calculations in super-Yang-Mills can be 
trusted in the region
\be\label{symm} 
g^2_{eff} \ll 1
  ~~~~\Rightarrow~~~~ \left\{ \ba{l}  U\gg (\gym^2 N)^{1/(3-p)}~,~~~~p<3
\\    U \ll 1/(\gym^2 N)^{1/(p-3)}~,~~~~p>3 \ea  \right. 
\ee

The type II supergravity solution describing 
 $N$ coincident extremal  Dp-branes
is (in the string frame)  \cite{hs}
\beq\label{bac}
&& ds^2=f_p^{-1/2}(-dt^2+dx_1^2+...+dx_p^2)+f_p^{1/2}
(dx_{p+1}^2+...+dx_9^2),\non
&& e^{-2(\phi-\phi_{\infty})}=f_p^{(p-3)/2},\\                
&& A_{0...p}=-\frac12 (f_p^{-1}-1)\nn,      
\eeq                                             
where $f_p$ is a harmonic function of the transverse coordinates 
$x_{p+1},...,x_9$
\be                                          
\alpha'^2 f_p= \alpha'^2 +\frac{d_p g^2_{YM} N }{U^{7-p}}~,~~~~~~~~~~~~~~
d_p = 2^{7-2p} \pi^{9-3p \over 2} \Gamma({7-p \over 2}) ~.
\ee 
In the field theory limit of eqs.(\ref{limit}), (\ref{limit2}) the solution
is
\beq\label{gsol}
&& ds^2=\al\left( \frac{U^{(7-p)/2}}{\gym\sqrt{d_p N}}dx^2_{||}+
\frac{\gym\sqrt{d_p N}}{U^{(7-p)/2}}dU^2+\gym\sqrt{d_p N}U^{(p-3)/2}
d\Omega^2_{8-p}\right) , \non
&& e^{\phi}=(2\pi)^{2-p} 
\gym^2\left( \frac{\gym^2 d_p N}{U^{7-p}}\right) ^{\frac{3-p}{4}} \sim
 \frac{\gef^{(7-p)/2}}{N}.
\eeq
Note that  the effective string coupling, $e^{\phi}$, is
 finite in the decoupling limit. 
In terms of $\gef$ the curvature
associated with the metric (\ref{gsol}) is 
\be\label{curg}
\al R\approx {1\over \gef } \sim \sqrt{{ U^{3-p} \over \gym^2 N }}.
\ee
      From 
the field theory point of view $U$ is an energy scale.
Thus, going to the UV in the field theory means taking the limit $U\r\infty$.
In this limit we see from eq.(\ref{gsol})  that for $p<3$ the
effective string coupling vanishes and the theory becomes UV free.
For 
 $p>3$ the coupling increases and we have to go to a dual description
before we can investigate it reliably.
This property of the supergravity solution is
 closely related to the fact that for $p>3$ the  super-Yang-Mills theories
 are 
non-renormalizable and hence, at short distances, new degrees of freedom
appear.

We  further note that, as in \cite{mal},
 we have $\al$ in front of the metric, which might lead to the 
incorrect conclusion that 
one should only
 consider the zero
modes of the fields.
However, a field theory quantum of energy $\omega$ will have proper energy 
$w_{proper}=\omega\sqrt{g^{tt}}={ 1\over \sqrt{\alpha'} }
\omega \frac{\gym\sqrt{d_n N}}{U^{(7-p)/2}}$
which remains  finite in string units. Therefore we can consider 
excitations which have proper enegies  comparable to the 
string mass. 

Consider the case that we break $U(N+1) \rightarrow U(N)\times U(1)$
by a Higgs expectation value $U$.  In the supergravity description
we get a Dp-brane sitting at the corresponding position $U$. 
The mass of a string stretched
between the $N$ branes and the probe is $m=U/2\pi$
 from the gauge theory point 
of view. This is a BPS state whose mass does not depend on the coupling. 
In the supergravity side this state is represented by  a string 
stretched between the probe and the horizon (at $r=0$). If we calculate its
 ``gauge theory'' energy from supergravity  we find that it is 
again $U$, since in curved
 space this energy contains a factor of $\sqrt{g_{rr}}$  
(due to the fact that we should consider the proper distance) which
is canceled by the 
$\sqrt{g_{tt}}$ factor needed to convert from local proper energies
to gauge theory energies (canonically conjugate to $t$). 

 
We will also consider near extremal configurations which correspond 
to the decoupled field theories at finite temperature. 
On the supergravity side we start from a near extremal black $p$-brane
solution and  we take
the limit (\ref{limit}) keeping the energy density on the brane finite. 
In this limit only the metric is modified 
\beq \label{nearextr}
&&ds^2=\al\left\{ \frac{U^{(7-p)/2}}{\gym\sqrt{d_p N}}
[ - ( 1 -{ U^{7-p}_0 \over U^{7-p} } )dt^2 + dy^2_{||}]+
\right.\non
&& \left.
\frac{\gym\sqrt{d_p N}}{U^{(7-p)/2} ( 1 - {U^{7-p}_0 \over U^{7-p}} )
}dU^2+\gym\sqrt{d_p N}U^{(p-3)/2}
d\Omega^2_{8-p}\right\}
\eeq
The dilaton is the same as in (\ref{gsol}) and 
\be \label{enerdens}
U_0^{7-p} = a_p \gym^4 \epsilon,\;\;\;\;\;\;\ a_p={ \Gamma({9-p \over 2} )
 2 ^{11-2p} \pi^{13-3p\over 2} \over (9-p) }
\ee
Here $\epsilon$ is the energy density of the brane above extremality and 
corresponds to 
the energy density of the Yang-Mills theory. With these formulas one
can calculate the entropy per unit volume 
and we find 
\be\label{a} 
s = {S \over V} = 
\left({ \Gamma({9-p \over 2})^2 2^{43-7p} \pi^{13-3p} 
\over (7-p)^{7-p} (9-p)^{9-p} } \right)^{1 \over 2 (7-p)}
 \gym^{ p-3 \over 7-p} \sqrt{N} 
\epsilon^{ 9-p \over
2(7-p) } 
\ee
The temperature follows from the first law of thermodynamics.

In order to trust the type II supergravity solution (\ref{gsol})
we need both the curvature (\ref{curg})  and  the dilaton
(\ref{gsol}) to be small. This implies
\beq \label{stringregion}
1 \ll  \gef^2 \ll N^{4 \over 7-p} ~ .
\eeq
We see that, as expected, the perturbative 
 super-Yang-Mills and supergravity
descriptions do not overlap (see eq.(\ref{symm})).
We will later see that the supergravity description can be extended
to  the region $ N^{4 \over 7-p}< g^2_{eff} $ but in terms of a dual 
theory. In particular, we see that we can trust the entropy 
computation (\ref{a}) as long as the energy density above extremality
is such that $g_{eff}(U_0)$ obeys (\ref{stringregion}) with
$U_0$ as in (\ref{enerdens})

The isometry group of the metric (\ref{gsol})
 is  $ISO(1,p)\times SO(9-p)$ (for $p\not =3$).
   From the super-Yang-Mills point of view  the $ISO(1,p)$ symmetry 
 is  the Poincare  symmetry and 
 $SO(9-p)$  is  the R-symmetry.
It is an R-symmetry since spinors on the world-volume of the Dp-branes 
 transform also as  spinors in  the  directions transverse
 to the
brane, and thus under  $SO(9-p)$,
 whereas the brane scalars
 transform in the vector representation of $SO(9-p)$.

\section{2+1 super-Yang-Mills and  D2-branes}

We start by considering a collection of $N$ 
D2-branes in the super-Yang-Mills limit,
\beq \label{decoup}
&& U=\frac{r}{\al}=\mbox{fixed},\;\;\; ~~~~ \gym^2=\frac{g_s}{\sqrt{\al}}
=\mbox{fixed},\;\;\; ~~~~~ \al\r 0,
\eeq
where $g_s= e^{\phi_{\infty}}$,  $g_{Y.M}$  is the Yang-Mills coupling
 constant which has dimensions of (energy)$^{1/2}$ and $U$ is the
expectation value of the Higgs. After taking the limit (\ref{decoup})
we decouple the bulk from the theory on the D2 branes which turns out
to be a   
U(N) super-Yang-Mills in 2+1 dimensions, with 16 supersymmetries. 
At a given energy scale, $U$, the dimensionless effective 
coupling  of the gauge theory is
$g^2_{eff} \sim \gym^2 N/U$ and, hence, perturbative 
super-Yang-Mills  can be trusted in the UV region
where $g_{eff}$ is small
\be
 \gym^2 N \ll U .
\ee
The supergravity  solution of $N$ D2-branes  \cite{hs} yields in this limit
\beq\label{sol}
&& ds^2=\al\left( \frac{U^{5/2}}{\gym\sqrt{6 \pi^2 N}}dx^2_{||}+
\frac{\gym\sqrt{6 \pi^2 N}}{U^{5/2}}dU^2+
\gym\sqrt{6 \pi^2 N/U}d\Omega^2_6\right) \non
&& e^{\phi}=\left( \frac{\gym^{10}6 \pi^2 N}{U^5}\right) ^{1/4}.
\eeq
The type II 
supergravity description can be trusted when the curvature (\ref{curg})
in string units 
and 
 the effective string coupling are small
\be
   g^2_{YM} N^{1/5}   \ll  U\ll \gym^2 N.
\ee
We see that a necessary condition is to have $N \gg 1$. 
In the region $g_{eff}\approx 1$ we have 
 a transition 
between the perturbative super-Yang-Mills description and the supergravity 
description. 

In the region $U< \gym^2 N^{1/5}$ the dilaton becomes large. In other
words the local value of the
radius of the eleventh dimension, $R_{11}(U)$, becomes larger than the
 Planck scale since $R_{11}=e^{2\phi /3}l_p$.  
Even though the string theory is becoming strongly coupled we will be able
to trust the supergravity solution if the curvature is small enough
in eleven dimensional Planck units.
The relation between  the eleven dimensional  metric and 
the ten dimensional type IIA string metric, dilaton and
gauge field is 
\be\label{1011}
ds^2_{11}=e^{4\phi /3 }(dx_{11}+A^{\mu}dx_{\mu})^2+
e^{-2\phi /3 }ds^2_{10}
\ee 
\begin{figure}
\begin{picture}(250,180)(0,0)
\vspace{-5mm}
\hspace{30mm}
\mbox{\epsfxsize=90mm \epsfbox{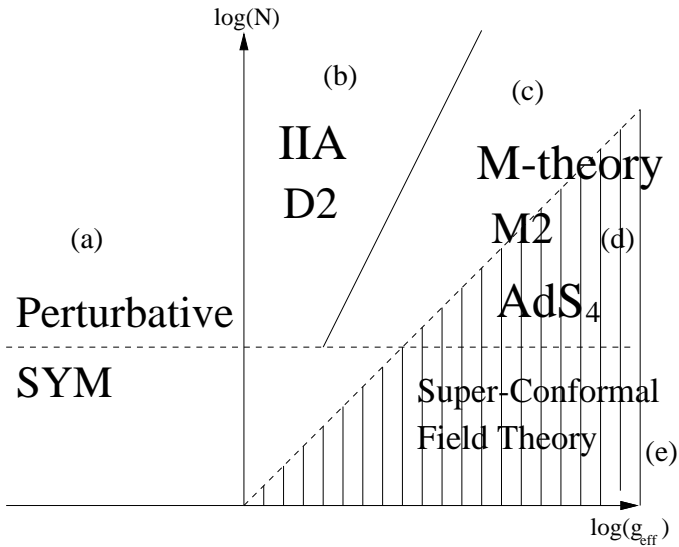}}
\end{picture}
\caption{The D2-brane map:
The horizontal dashed line separates between the small $N$ region and the
large $N$ region. The  other dashed line separates  the IR
region from the rest.
The UV description is via perturbative super-Yang-Mills (a). 
In  the IR the theory flows to a super-conformal region
 (the  marked region) with SO(8) R-symmetry \cite{sei}.
For large $N$ we have a
region described by IIA supergravity (b), a region described by 
the periodic array of M2-branes  solution in 
eleven dimensions (c) and finally in the IR we have M-theory on the  
$AdS_4\times S^7$ background (d).}
\end{figure}
which implies that the curvature in 11D Planck units
is
\be
l^2_p R \sim e^{2\phi /3}\frac{1}{g_{eff}}\sim 
\frac{1}{N^{1/3}}\left( \frac{g^2_{YM}}{U}\right) ^{1/3}.
\ee
For large $N$, in  the region $\gym^2<U $, the curvature in 11D
 Planck units is
small.
We show below that in the region
  $U<\gym^2$ we should use a different solution 
corresponding to M2-branes  localized on the circle associated with the 11th
dimension.  
 The  curvature of $N$ M2-branes in the field theory limit 
is $l^2_p R \sim 
\frac{1}{N^{1/3}}$ \cite{mal}.
Note that the curvature does not depend on $U$ since 
the
theory is conformal in the IR limit.
We conclude, therefore, that for large $N$ the supergravity description
 is valid in the region $U\ll \gym^2 N^{1/5}$. 

The 11 dimensional solution that we get by uplifting the D2 brane solution 
 (\ref{sol}) using eq.(\ref{1011}) is not 
exactly the M2 brane solution\footnote{``Uplifting'' means that we
 find the eleven
dimensional metric and fourform field strength (independent of $x^{11}$)
which give the solution (\ref{sol}) upon Kaluza-Klein reduction.}. 
The uplifted solution is the M2 brane solution averaged over one of  the
transverse directions. 
Let us be more explicit.
  The M2 brane solution is characterized by an harmonic function $H$
and is given by 
\begin{equation} \label{mtsol}
ds^2_{11} = H^{-2/3} dx^2_{||}  + H^{1/3} dx^2_{\perp}
\end{equation}
and there is also a fourform field strength given in terms of $H$. 
When we take $H \sim   N/ x^6$ we have a solution where the 
M2 branes are localized in the eight transverse non-compact 
dimensions. If one of the dimensions is compact 
(let us say the 11th dimension)
we can take $H  \sim N/ r^5$ where now $r$ denotes the radial distance in 
the seven transverse 
non-compact dimensions. This is the solution we get from
uplifting (\ref{sol}).
We will later see that this solution is unstable when we raise the
 temperature a  little bit.
 The more physical solution is the one in which we take the
M2 branes
to be localized in the compact dimension so that the harmonic function is 
\be\label{superp}
H = \sum_{ n = - \infty}^{\infty}  { 2^5 \pi^2 
  N l_p^6 \over ( r^2 + (x_{11} -x_{11}^0 + 2
  \pi n
 R_{11})^2)^3}
\ee
with $x_{11} \sim x_{11} + 2 \pi R_{11} $.
For distances much larger than $R_{11}$ we
can Poisson resum this expression to 
\be \label{resumed}
l_p^3 H = {6 \pi^2 \gym^2 N \over U^5 } + 
\sum_{m = 1 }^{\infty}  N e^{ - m U/\gym^2
  }\cos(mx_{11}^0/R_{11}) {\cal O}(U^{-5})
\ee
where we have used that $R_{11} = g^{2}_{YM} {\alpha'} $.
For  $ \gym^2 \ll U$ 
we can, therefore, use the uplifted solution to describe the physics while for
 smaller values of
$U$ we should use (\ref{superp}).
 Note that for such small energy
scales it becomes 
necessary to specify the expectation value of $\phi^{11} = \gym^2 
x_{11}^0/R_{11} $ which is the
new scalar coming
from dualizing the vector in 2+1 dimensions.
In fact it was shown in \cite{instd2} that the $v^4$ term in the 
effective action of a D2-brane probe receives instanton
contributions which produce the whole series (\ref{resumed}).  
For very low  energies 
\be\label{ir}
U \ll \gym^2,
\ee
we are very close to the M2 branes and we can neglect the ``images''
in (\ref{superp}).
Thus,  the solution will resemble that of M2 branes in non-compact space
and we have the conformal field theory with SO(8) symmetry, which is
the case described in \cite{mal}.  
We note that the physical size of the eleventh circle at the point of
the transition 
 between the localized and the delocalized solution, $U \sim \gym^2$,
 is much larger than the Planck
length, $ R_{11}^{phys}|_{U=g_{YM}} \sim l_p  N^{1/6} $. 
Hence,  we can trust the supergravity
solution whenever we have some non-trivial dependence of the solution
on $x_{11}$ and we have a smooth transition to the IIA supergravity 
regime when the physical size of $R_{11}$ becomes small. In other
words, starting from the IR and flowing to the UV
the eleven dimensional supergravity solution becomes independent
of $x_{11}$ before $R_{11}^{phys}(U) $ becomes smaller than the eleven 
dimensional Planck length.

\subsection{Near extremality}\label{nD2}

We would like to consider now finite temperature configurations in the 
super-Yang-Mills  theory. 
We always take the energy above extremality and the temperature to be
finite in the decoupling limit (\ref{decoup}). 
The supergravity solution  corresponding to a near extremal D2 brane 
in IIA string theory has one more harmonic 
function $h = 1 - U_0^5/U^5$ 
as in (\ref{nearextr}).
 The parameter $U_0$ is finite in the decoupling limit and is given by
\be
U^5_0  =  {240 \pi^4 \over 7} \gym^4  \epsilon,
\ee
where $\epsilon $ is the energy density. This is the energy density 
of the field theory, and
it corresponds to the energy density of the brane above extremality. 
This solution describes the physics appropriately as long as 
$U_0  \gg \gym^2$ (although it might be necessary to uplift the solution
to eleven dimensions). When $U_0 \ll \gym^2$ the uplifted solution
(which is  a valid solution of the equations of motion)  
becomes unstable \cite{instability}. This is just a classical 
instability associated
with non-extremal black p-branes. 
There is an easy way to understand it. 
Let us first remember how the delocalized solution is generated. 
We start with M-theory on $T^2\times S^1$ and we take the
radius of $S^1$ to be much larger than the Planck length. 
We consider a Schwarzschild black hole solution in 
the 7+1 non-compact dimensions. 
Then we apply   a 
boost along one of the directions of the $T^2$ (a symmetry of
the supergravity equations) which generates some Kaluza-Klein
momentum charge and then  we U-dualize  it into M2 branes wrapped
on $T^2$. This procedure gives a solution which does not
 depend on the coordinate  along 
$S^1$, which we call the ``eleventh'' dimension.
This  can actually be done for {\it any} supergravity solution 
of M-theory on $T^2$, regardless of whether
it  is localized
on the extra $S^1$ or not.   
More explicitly, if we start from an uncharged static solution in 
8+1 dimensions 
 $ds^2_{1+8} = g_{00} dt^2 + g_{i j}  
d x^i  d x^j $ (where $g_{00},~g_{ij}$ depend on $x_i$),
then the  solution with M2 brane charge obtained after performing this 
process, ``uplifted'' to eleven dimensions  will be 
\be
ds^2_{11 } = ( \cosh^2 \alpha + \sinh^2 \alpha g_{00} )^{-2/3} [g_{00}dt^2 
 + dy_1^2 + dy_2^2]+ 
( \cosh^2 \alpha + \sinh^2 \alpha g_{00} )^{1/3} g_{ij}dx^i dx^j 
\ee
where $y_i$ are the two spatial coordinates along the brane and the
periodic
coordinate is among the $x_i$.  
Therefore, properties of the uncharged solution will translate into properties
of the near extremal M2 brane configuration.
Consider the uncharged 
Schwarzschild solution in 7+1 non-compact dimensions  described above.
It is translational invariant along $S^1$. It is, therefore, a black string.
 This black string is unstable when the Schwarzschild radius of the
string becomes smaller than the radius of $S^1$ \cite{instability}. 
%
It seems plausible to think that the 
solution  decays into a solution which is localized along the circle, which
looks like a black hole in 8+1 dimensions (though we have not
shown this explicitly)\cite{instability}. 
This is supported by the observation that the entropy
of the 8+1 black holes is bigger than the entropy of the black string
when their Schwarzschild radii are smaller than the radius of the circle. 
Of course, in order for the supergravity analysis to be valid all  these 
 radii should be 
much bigger than the Planck length. 
To describe more precisely this transition  
one should find supergravity solutions which are
localized in the compact dimensions.
 These solutions would look like 
an infinite array
of Schwarzschild black holes (they do not  collapse on each 
other because of the periodicity conditions).
 Some solutions of this type were found 
in 3+1 dimensions with one compact spatial dimension
\cite{rmyres}. 
After we do the boosts and U-duality transformations to produce M2
 brane charge we
see that the point where the Schwarzschild radius is comparable to 
the radius of the $S^1$ corresponds,
in the new solution, to the point where 
\be \label{trans}
U_0 \sim \gym^2 \ .
\ee 
In summary, for $U_0 > \gym^2 $ we have a translational invariant 
solution (along $x_{11}$)  while for 
$U_0 < \gym^2$ we have a localized solution. 
When $U_0 \ll \gym^2$ we can find an approximate near extremal 
solution by considering a linear superposition of Schwarzschild
black holes and performing the above procedure. The thermodynamics
for  this solution will be, by construction, the same as the one
for the near extremal M2 brane in non-compact 11 dimensions. This
is what we expect for 2+1 SYM at low energy, i.e. that 
 the results should be
those of the corresponding IR superconformal field theory.


\bl{\em Conclusions:}

\bl We are always considering 2+1 dimensional super-Yang-Mills.
This theory has a large N dual which is 
the supergravity solution described above. 
This supergravity solution has various regions. The supergravity 
description requires that $U\ll \gym^2 N$ (when 
$ \gym^2 N \ll U $ perturbative
Yang-Mills is a good description).
In the region  $N^{1/5}<U/g^2_{YM}<N$ we have a IIA string theory
description, we expect to have strings, etc. as in \cite{mal}.
When $U/g^2_{YM} < N^{1/5} $ we should use an eleven dimensional 
supergravity solution.
 The transition from the ten dimensional 
to the eleven dimensional solution is smooth from the point of
view of supergravity and the gradient of the dilaton remains always small
 (for large $N$). When $U \ll g^2_{YM}$ the geometry becomes
that of $AdS_4\times S^7$ which is the one that corresponds to 
the low energy conformal field theory with 
SO(8) R-symmetry.  


Similarly when we consider a near extremal configuration we see that 
for very low temperatures the behavior is that of the M2
 brane conformal field 
theory. One could, in principle, follow the transition 
between the near extremal M2 brane
and the near extremal D2 brane behavior if one knew more precisely 
the localized supergravity
solution. 
 
\section{1+1 super-Yang-Mills and D1-branes}

Next we turn our attention to a collection of $N$ D1-branes 
corresponding to the case $p=1$ in section 2.
The decoupling  limit  takes now the form
\be\label{limD1}
U=\frac{r}{\al}=\mbox{fixed},
\;\;\;\gym^2={1 \over 2 \pi} \frac{g_s}{\al}=\mbox{fixed},
\;\;\;\al\r 0.
\ee
The supergravity solution in this limit yields 
\beq\label{solD1}
&& ds^2=\al\left( \frac{U^{3}}{\gym\sqrt{2^6\pi^3N}}dx^2_{||}+
\frac{\gym\sqrt{2^6\pi^3N}}{U^{3}}dU^2+\gym\frac{\sqrt{2^6\pi^3N}}{U}
d\Omega^2_6\right), \non
&& e^{\phi}=\left( \frac{\gym^6 2^8\pi^5N}{U^6}\right) ^{1/2}.
\eeq
Super-Yang-Mills  in $1+1$ dimensions  is super-renormalizable
and  it  can be trusted at high energies
$\gym\sqrt{N}\ll U$.
The curvature in string units 
(\ref{curg})
is  small for $U\ll\gym \sqrt{N}$. From 
eq.(\ref{solD1}) we see that the expansion in string coupling is valid 
in the region $ \gym N^{1/6}\ll U$.
Therefore, the type IIB supergravity solution (\ref{solD1}) 
can be trusted in the region 
$ \gym N^{1/6} \ll U \ll \gym \sqrt{N}$.
In the region $U\ll g_{YM} N^{1/6}$ the string coupling is large.
To get a more accurate description we need to 
apply S-duality, which 
 takes $\phi\r -\phi$ and hence $g_s\r \tilde{g_s}=1/g_s$.
Since the ten dimensional Newton constant, $G_N^{10}=8\pi^6g_s^2\al^4$,
 is invariant
S-duality also takes $\al\r\tilde{\a}^{'}=g_s\al$.
Note that in the decoupling limit we considered $\tilde{\a}^{'}\r 0$.
S-duality, therefore, maps (\ref{solD1})  to the small $r$  region of the 
fundamental string solution
\cite{dhar}
\begin{figure}
\begin{picture}(250,180)(0,0)
\hspace{18mm}
\vspace{-5mm}
\mbox{\epsfxsize=90mm \epsfbox{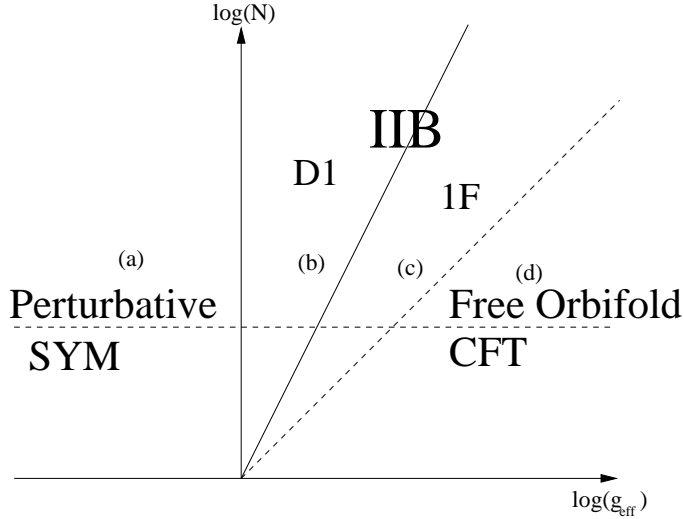}}
\end{picture}
\caption{The D1-brane map:
The horizontal dashed line separates between the small $N$ region and the
large $N$ region. The  other dashed line separates between the IR
region and the rest.
For any  $N$ the UV region is described by perturbative SYM (a)
and in the IR
 by a  free orbifold CFT  (d).
When $N$ is large there is an intermediate region (b,c) which 
is described by a IIB supergravity solution interpolating between 
 the D1 brane solution (b) and the 
F-string solution (c). }
\end{figure}
\beq
&& ds^2=\tilde{\a}^{'}\left( \frac{U^6}{\gym^42^7 \pi^4 N}dx^2_{||}+
\frac{1}{2 \pi \gym^2}dU^2+\frac{U^2}{2 \pi \gym^2}d\Omega^2\right), \non
&& e^{\phi}=\left( \frac{\gym^62^8 \pi^5 N}{U^6}\right) ^{-1/2}.
\eeq 
In the IR limit ($U\r 0$) the string coupling vanishes.
The curvature in the new string units is
\be\label{24}
\tilde{\a}^{'} R \sim \frac{\gym^2}{U^2},
\ee
and  does not depend on $N$.
The reason is that the  dependence of the metric on $N$ drops out 
after the coordinate change $x_{||}\r x_{||}/\sqrt{N}$.
Note that
 there is a curvature singularity in the IR limit.
This means that the supergravity description
 breaks down\footnote{ This breakdown of the fundamental string
solution was also studied by Sen \cite{senbh}.}
 for small $U$.
In fact the IR limit of super-Yang-Mills is a trivial orbifold 
($(R^8)^N/S_N$)
conformal field theory.
Furthermore, the first irrelevant operator that appears in this theory
was found in \cite{dvv}. We could now consider the theory
at finite temperature and ask when the orbifold CFT is a valid
description. The orbifold is characterized by $N$ fields and the
first correction is given by the twist   operator  
 ${1\over g_{YM} } V_{ij}$ where $i,j$ label
the two fields on which the twist is acting \cite{dvv}.
 The power of $g_{YM}$
 follows from dimensional analysis. 
We compute the partition function for the orbifold and we 
see at which temperature the correction due 
to the twist operator becomes large. We find that 
the free field theory
will be a good approximation if the temperature satisfies 
$T\ll g_{YM}/N^{1/2}$. This in turn translates into a parameter $U_0$ of
the near extremal solution (\ref{nearextr}) which is $U_0 \ll g_{YM}$
\footnote{The same conclusion was reached in \cite{dvv2}.}. 
It implies that the point where the supergravity solution breaks
down is related to the point where the free conformal field theory
takes over. Notice that it would have been impossible to have 
a free field theory   dual to a supergravity system.

\bl {\em Conclusions:}

\bl The 1+1 dimensional super-Yang-Mills under consideration flows 
for any $N$
both in the  UV and IR  to  
free field theories.
The large $N$ dual of these theories is a supergravity 
solution that is valid in the intermediate
region $ g_{YM} \ll U \ll g_{YM} \sqrt{N}$.
At both ends of this limit the curvature grows and the solution breaks
 down.
Furthermore for 
$ g_{YM} \ll U \ll g_{YM} N^{1/6}$ the proper description is through
the fundamental string solution while for $  g_{YM} N^{1/6}
 \ll U \ll  g_{YM} \sqrt{N}$ we should use the D-string supergravity
solution.

\section{D0-branes and super quantum mechanics}

In this section  we consider the super quantum mechanical theory
associated with $N$ D0-branes in the  limit
\beq\label{sqm}
&& U=\frac{r}{\al}=\mbox{fixed},\;\;\gym^2=\frac{1}{4\pi^2}
\frac{g_s}{\al^{3/2}}=\mbox{fixed},\;\;\;\al\r 0.
\eeq
Notice that $U/\gym^{2/3} \sim r/l_p$ as  in \cite{kpdkps}.
Perturbation theory in the quantum mechanics can be trusted 
at high energies $\gef\ll 1$ which gives
 $U>\gym^{2/3}N^{1/3}$.

The supergravity solution in the decoupling limit gives 
\beq\label{b}
&& ds^2=\al\left( -\frac{ U^{7/2}}{ 4\pi^2\gym\sqrt{15\pi N}}dt^2+
\frac{4\pi^2\gym\sqrt{15\pi N}}{U^{7/2}}d U^2+
{4\pi^2\gym\sqrt{15\pi N}\over U^{3/2}} d\Omega^2\right), \non
&& e^{\phi}= 4 \pi^2 \gym^2 \left( \frac{240 \pi^5 \gym^2N}{U^7}\right)
 ^{3/4}.
\eeq
For this solution the effective string coupling and the curvature in string
 units are small in the region
\be
 \gym^{2/3}N^{1/7}\ll U \ll \gym^{2/3}N^{1/3}
\ee
Thus, in this region one can trust the type IIA description.


\begin{figure}
\begin{picture}(250,180)(0,0)
\vspace{-5mm}
\hspace{30mm}
\mbox{\epsfxsize=90mm \epsfbox{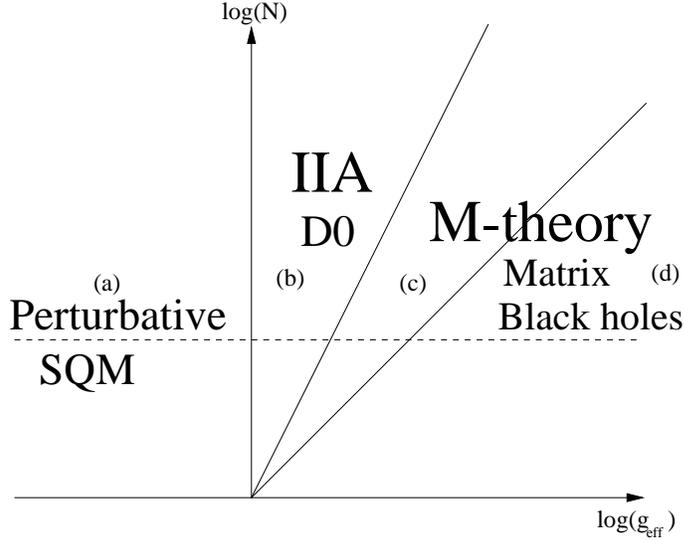}}
\end{picture}
\caption{The D0-brane map:
The horizontal dashed line separates between the small $N$ region and the
large $N$ region. 
For any  $N$ the UV description is via perturbation theory in
super quantum mechanics (a).
For large $N$ we have a region (b) which is described by the
 IIA D0 brane solution,
which for smaller energies becomes a gravitational wave background in eleven
dimensions (c).
Finally, at very  low energies (d) we enter into the matrix black hole
region.}
\end{figure}

Now we would like to study the low energy region. 
%
To this end we need to uplift the solution to eleven dimension.
This solution can be generated by starting with an uncharged black string
along $x_{11}$ and then boosting it 
along $x_{11}$ while taking the limit 
\be
\gamma\r \infty,\;\;\; \gamma \mu=
\frac{N }{2\pi R_{11}^2}=\mbox{fixed},
\ee
where 
 $\mu$ is the
mass per unit length of the black string in its rest frame.
As we show below this plane wave description cannot be trusted 
for $U<\gym^{2/3}
N^{1/9}$.
This region is
closely related to the matrix
model black holes \cite{bhmatrix,homa}. 

\subsection{Matrix black holes}

The  eleven dimensional 
 solution which corresponds to the uplifted near extremal
D0 brane solution
is translational  invariant along the circle.
This solution is unstable for small enough energies above extremality.
In the D2-M2 case we estimated the energy scale at which this instability 
was happening by
starting from the neutral black string solution and tracing through
the steps in the solution generating technique. In this case we do the
same. The solution is gotten by performing a boost along an uncharged 
black string. As explained in \cite{homa} the instability appears
when 
\be
U_0 \sim \gym^{2/3} N^{1/9}
\ee
which is the ``correspondence'' point of \cite{bhmatrix,homa}.
Of course this localization makes sense only in the large $N$ limit where
we keep a fraction $M\ll N$ of the total momentum as gravitons
so that the total system of gravitons plus black hole is in a momentum
eigenstate.

For $U_0 \ll  \gym^{2/3} N^{1/9}$ we have a Schwarschild black hole boosted
along the eleventh direction. It is interesting to note that the
Schwarschild radius of the black hole in Planck units is given by 
\be 
\left( \frac{r_s}{l_p}\right)^8 \sim { E^{1/2} N^{1/2}
\over \gym^{1/3} } \sim { U_0^{7/2} N^{1/2} \over \gym^{7/3} }
\ee
So that we trust the  gravity description if $U_0 \gg \gym^{2/3} N^{-1/7}$.
For lower values of the energy we expect that the system should start to 
behave more as a single graviton. Notice that if we keep the ratio
$r_s/l_p$ fixed then the energy above extremality goes as $E\sim 1/N$ as 
we expect \cite{bfss}.

\section{D4-brane, 4+1 SYM and the (0,2) 6-d SCFT on a circle}

Let us  consider now  a system of $N$ D4 branes 
 in the limit
\beq\label{12}
&& U=\frac{r}{\al}=\mbox{fixed},\;\;~~~\gym^2=(2\pi)^2g_s
\sqrt{\al}=\mbox{fixed},\;\;\;~~~
\al\r 0.
\eeq
This system is better described by considering 
a system of M5-branes
wrapped on the eleventh dimensional circle in M-theory and taking 
the limit $l_p \to 0$ while keeping 
$R_{11}= g_s\sqrt{\alpha'}=  g^2_{YM}/(2\pi)^2$ fixed.
So  we have  the (0,2) six dimensional conformal field theory on a circle, 
which, at low energies, reduces to 4+1 dimensional super-Yang-Mills 
\cite{z2roz}. 

The supergravity description involves a type IIA supergravity 
region and an M-theory region. The IIA solution is 
\beq
&& ds^2=\al\left( \frac{2 \sqrt{\pi} U^{3/2}}{ \gym\sqrt{N}}dx^2_{||}+
\frac{\gym\sqrt{N}}{2 \sqrt{\pi}U^{3/2}}d U^2+
{\gym\sqrt{N U} \over 2 \sqrt{\pi}} d\Omega^2\right), \non
&& e^{\phi}=\left( \frac{ U^3\gym^6}{2^{10} \pi^9N}\right) ^{1/4}.
\eeq
The ``perturbative'' super-Yang-Mills\footnote{
 We have put ``perturbative'' in quotes
because the theory is non-renormalizable. In principle we could
use perturbation theory to calculate diagrams which are finite. 
Examples of finite diagrams  are the $v^4$ terms \cite{bfss} (and, of 
course, all tree level diagrams).}
description 
 can be trusted in the IR region where the effective coupling is small,
$g_{eff}^2=\gym^2 N U\ll 1$. 
In the region where $ N^{-1} \ll g^2_{YM} U $ the curvature (\ref{curg}) is
 small in string units.
The dilaton is small for $ g^2_{YM}U \ll N^{1/3}$.
Therefore the  type IIA supergravity solution can be trusted in the region
$ N^{-1} \ll g^2_{YM} U\ll N^{1/3}$. 
\begin{figure}
\begin{picture}(250,160)(0,0)
\vspace{-5mm}
\hspace{30mm}
\mbox{\epsfxsize=90mm \epsfbox{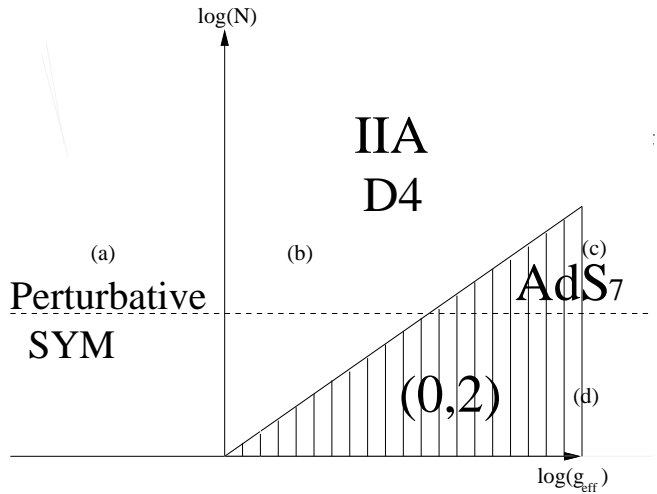}}
\end{picture}
\caption{The D4-brane map:
The horizontal dashed line separates between the small $N$ region and the
large $N$ region. 
The UV region is described by a super-conformal
theory on a circle  (the marked region) (c,d),
 which is dual for large $N$ to 
  M-theory on a background $AdS_4\times S^7$ (c) (with an 
identification). 
In  the IR the theory is described by ``perturbative''  super-Yang-Mills (a).
For large $N$ we have the intermediate 
  region (b) described by the IIA D4 brane solution.}
\end{figure}
In the region $ N^{1/3}\ll g^2_{YM} U $ the dilaton is large and the
description is via 11D supergravity.
Using eq.(\ref{1011}) we find the 11D solution of $N$ NS5-branes wrapped
along the $x_{11}$ direction,
\be\label{02}
ds^2=l_p^2\left( \frac{\tilde{U}^2}{(\pi N)^{1/3}}dx_{||}^2+4
(\pi N)^{2/3}
\frac{d\tilde{U}^2}{\tilde{U}^2}
+(\pi N)^{2/3}d\Omega_4^2\right),
\ee
where $\tilde{U}^2= (2\pi)^2 U/g^2_{YM} = U/R_{11}$.
This describes a M-theory background of $AdS_7\times S^4$ \cite{mal}
with an identification along a circle. 
The radius of the Anti-de Sitter space and the radius of the sphere
are large (in eleven dimensional Planck units) for large $N$.
 Hence,
the solution can be trusted for large $N$, as long as the physical 
length of the circle that we are identifying is large enough. 

Note that the $\gym$ dependence drops out in (\ref{02}).
This is a result of the theory being conformal at the UV and $\gym^2$
having dimensions of length.

\bl {\em Conclusions:}

\bl In the present case we are 
 dealing with the (0,2) theory compactified on a circle, which
becomes 4+1 dimensional Yang-Mills at low energies. This theory is free
in the IR. 
For large $N$ we have a dual supergravity description. It involves a type
IIA supergravity solution in the region $N^{-1} \ll g^2_{YM} U \ll
N^{1/3}$ and  M-theory on $AdS_7\times
S^4$ with identifications 
in the region $ N^{1/3} \ll \gym^2 U$. 
%

\section{D5-branes and IIB NS fivebranes}  

For the system of $N$ D5-branes the 
 relevant decoupling  limit is
\be\label{limitD5}
U=\frac{r}{\al}=\mbox{fixed},
\;\;\;\gym^2=(2 \pi)^3 g_s\al=\mbox{fixed},
\;\;\;\al\r 0.
\ee
In this limit the supergravity  solution gives
\beq\label{solD5}
&& ds^2=\al\left( \frac{(2\pi)^{3/2} U}{\gym \sqrt{ N}}dx^2_{||}+
\frac{\gym \sqrt{ N}}{(2\pi)^{3/2}U}dU^2+
{ \gym \sqrt{ N}U \over (2\pi)^{3/2}} d\Omega^2_6\right), \non
&& e^{\phi}= \frac{\gym U}{(2\pi)^{3/2} \sqrt{N}}.
\eeq
The super-Yang-Mills can be trusted in  the 
IR region $\gym U\ll\frac{1}{\sqrt{N}}$.
In the region $\frac{1}{\sqrt{ N}}\ll \gym U\ll\sqrt{N}$ the
string coupling and the curvature in string units are
 small so one can trust the 
type D5-brane supergravity  solution.
For $\sqrt{N}\ll \gym U$ the string coupling is large and we have to
go to the S-dual system of $N$ NS5-branes where
\be
\tilde{\a}^{'}=g_s\al=\gym^2/(2\pi)^3
\ee
and, therefore, remains finite in the limit (\ref{limitD5}).
\begin{figure}
\begin{picture}(250,180)(0,0)
\vspace{-5mm}
\hspace{18mm}
\mbox{\epsfxsize=90mm \epsfbox{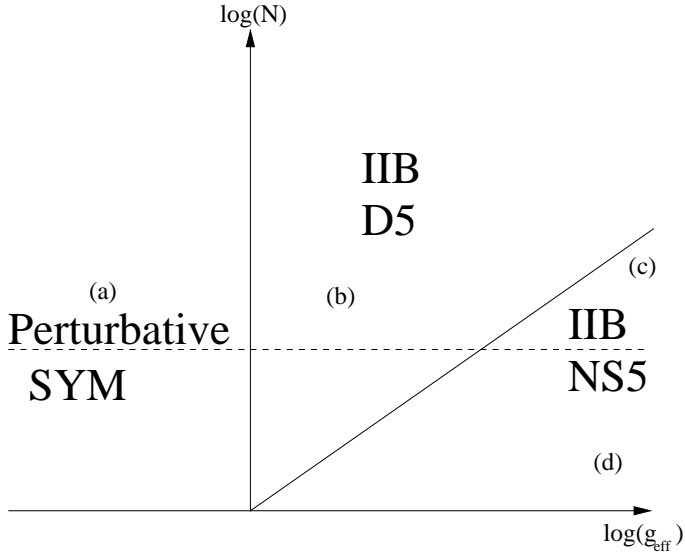}}
\end{picture}
\caption{The D5-brane map:
The horizontal dashed line separates between the small $N$ region and the
large $N$ region. 
In the IR the theory  are described by ``perturbative'' SYM (a).
For  the UV description is via type IIB on NS background (c,d) which
has supergravity description for large $N$ (c).
For large $N$ there is an intermediate region described by 
the D5 brane background (b).}
\end{figure}
The solution for the NS5-branes is
\beq 
&&ds^2= dx^2_{||} + \tilde{\a}^{'}\left( 
 \frac{N}{U^2}dU^2+Nd\Omega^2\right),\non
&&  e^{\phi}=\left( \frac{ (2\pi)^{3} N}{\gym^2 U^2}\right) ^{1/2}.
\label{nsfive} \eeq
The curvature  in string units is $ \tilde{\a}^{'}R \sim \frac{1}{N}$,  
and, therefore,  it is small in string units  for large $N$ and hence the
supergravity  description is valid for large $N$ \cite{mal2}.
It is, however, possible to have an exact conformal field theory 
description of this background as a classical solution of string
theory (which is appropriate in this region where the dilaton 
is becoming small) \cite{wssol}.
This case is different from other field theory cases 
because massive particles
can propagate all the way to infinity along the throat.
The throat is infinite in the limit (\ref{limitD5}).

\bl {\em Conclusions:}

\bl We have the  theory  defined as the $g_s \to 0$ limit of 
NS IIB fivebranes \cite{seisix}.
 This  theory is characterized by a scale 
$\gym^2 \sim \tilde \alpha' $.
 For large $N$ we can analyze properties of the
theory using supergravity. 
This  theory  flows in the IR, $\gym U \ll N^{-1/2} $, to super-Yang-Mills.  
For scales $ N^{-1/2} \ll \gym U  \ll N^{1/2} $ we should use the D5 brane
supergravity solution.
 Finally for $ N^{1/2} \ll \gym U$ we should
use the NS fivebrane solution. This solution involves an infinite 
 throat region
(for large $U$) which does not decouple from the physics \cite{mal2}
and should therefore be included in the description.



\section{IIA NS fivebranes}

This system was studied in \cite{mal2} for high temperatures. 
In this section we add some comments which are relevant for the 
description at lower temperatures (and large $N$). 
We observe  that in the region where the dilaton 
is large we can go to an eleven dimensional description 
where nothing singular happens (for large $N$), and we
flow in the IR to $AdS_7\times S^4$ which is the large $N$
dual of the (0,2) conformal field theory, so that things work
as expected. 
We are interested in the   system of $N$ NS fivebranes in the limit
\be
g_s \to 0 ,  \, \, \,   \alpha' = {\rm fixed},  \, \, \,  U =
 { r \over g_s \alpha' } = {\rm fixed}
\ee
Notice the additional factor of $g_s$ in the definition of $U$, this 
ensures that the tension of D2-branes stretched between different
fivebranes
is constant. 
The supergravity solution is the same as (\ref{nsfive}).
This system was analyzed in \cite{mal2} for large temperature. 
For low energies it is more appropriate to go to eleven dimensions and,
therefore, consider
 M5 branes transverse to a compact circle. 
An instability of the type described for the M2 branes will lead us
to consider
M5 branes localized on the circle. (The instability would be 
present only for non-zero temperature). 
 The supergravity solution is
then
\beq\label{multim5}
&& ds^2 = H^{-1/3} dx^2_{||}  + H^{2/3} dx_{\perp}
\non
&& H = \sum_{n = -\infty}^{\infty} {\pi  N l_p^3 \over ( r^2 + (x_{11} + 2
 \pi R_{11} n)^2 
)^{3/2} } \sim { N \alpha' \over r^2 } +  N \sum_{m \geq 1} 
e^{- m r/R_{11} }
 {\cal O}( r^{-2})
\eeq
where $r/R_{11} = {U  \sqrt{\alpha'}} $.
Hence, for $U \ll {1 \over \sqrt{\alpha'} }$ we are very close to one of the
centers in (\ref{multim5}) and we flow into the 
(0,2) conformal field theory. 
Again as in the case of the D2 brane we see that we first go  into 
the eleven dimensional 
description and then into the localized description (if $N$ is
 sufficiently large). 
These localized solutions were analyzed in detail in \cite{ghmns}, but
 for the present discussion
it is very important that $N$ is large, otherwise we cannot trust the
 supergravity solutions.
Notice that even though the radius of the eleventh dimension seems to
 go to zero 
very far away from the NS brane, it becomes large close to the NS
brane.
It is so large that 
we are arguing that we should use a supergravity solution that
explicitly depends on 
the eleventh  dimension. 
If we heat up the system, considering it at a finite temperature, we
 expect to have a 
localized solution for small energy densities ($\epsilon \ll 1/\alpha'^3$), 
with the thermodynamic
behavior of the (0,2) theory.
An approximate solution can be found as for the D2-M2 brane case. For 
 larger energy densities  we expect to have a solution that is
translational
 invariant along the 
eleventh dimension. 

\section{D6-brane and 6+1 super-Yang-Mills}

The last system we analyze in this paper is the system on $N$ D6-branes.
The candidate decoupling limit in this case seems to be 
\beq\label{limD6}
&& U=\frac{r}{\al}=\mbox{fixed},\;\;~~~~
\gym^2=(2\pi)^4 g_s\al^{3/2}=
\mbox{fixed},\;\;\;~~~~
\al\r 0.
\eeq
which is better analyzed by going to M-theory on a circle with $N$
 Kaluza-Klein monopoles. The limit is now 
 simply $l_p = \mbox{fixed}$, $ R_{11} \to \infty$ which leaves 
an ALE singularity with 
$l_p = \gym^{2/3}/(2\pi)^{4/3}$.
The IIA supergravity solution in the limit (\ref{limD6}) gives
\beq\label{solD6}
&& ds^2=\al\left( \frac{(2\pi)^2 }{\gym}\sqrt{\frac{2 U}{ N}}dx^2_{||}+
{\gym \over (2\pi)^2 }
\sqrt{\frac{N}{2 U}}dU^2+{\gym \over (2\pi)^2 \sqrt{2}}
\sqrt{N}U^{3/2}d\Omega^2\right), \non
&& e^{\phi}= { g^2_{YM} \over 2 \pi} 
\left(2  \frac{U}{\gym^2N}\right) ^{3/4}.
\eeq
\begin{figure}
\begin{picture}(250,180)(0,0)
\vspace{-5mm}
\hspace{30mm}
\mbox{\epsfxsize=90mm \epsfbox{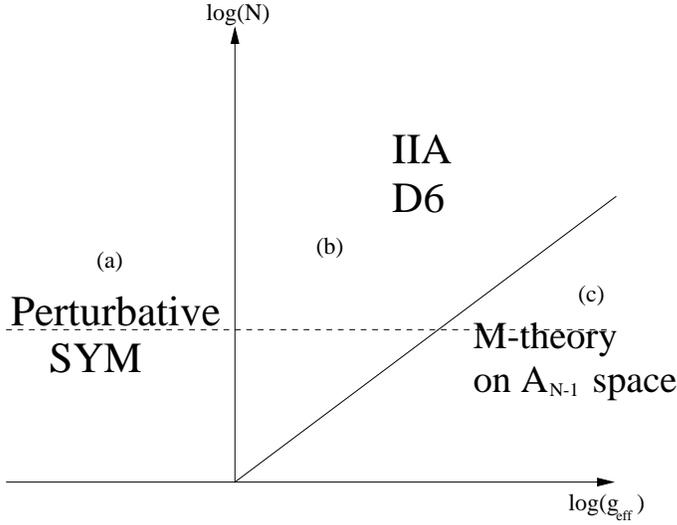}}
\end{picture}
\caption{The D6-brane map:
The horizontal dashed line separates between the small $N$ region and the
large $N$ region. 
The IR is described by SYM (a).
For any  $N$ the UV region is described by M-theory on an ALE space
with $A_{N-1}$ singularity (c).
In the large $N$ limit there is a region 
described by the IIA D6 brane solution (b).
}
\end{figure}
The super-Yang-Mills effective coupling is $g^2_{eff}=\gym^2NU^3$.
 Thus,
 super-Yang-Mills is a good approximation at low energies, 
$U\ll\frac{1}{\gym^{2/3}N^{1/3}}$.
The curvature 
(\ref{curg})
is small in string units in the region
$ \frac{1}{\gym^{2/3}N^{1/3}} \ll U $.
The dilaton is small in the region $U\ll \frac{N}{\gym^{2/3}}$.
Thus the type IIA supergravity solution can be trusted  in the region
$\frac{1}{\gym^{2/3}N^{1/3}}\ll U\ll \frac{N}{\gym^{2/3}}$.
In  the region $ \frac{N}{\gym^{2/3}}\ll U $ we get from eq.(\ref{solD6})
that $R_{11}^{phys}(U) \gg l_p$ hence  we should lift up the   type
IIA solution to eleven dimensions.
Using eq.(\ref{1011}) we get for the 11D metric
\be\label{kkmonshor}
ds^2=dx^2_{||}+\frac{l_p^3N}{2 U}dU^2+{ l_p^3NU \over 2} (d\theta^2+
\sin^2 \theta d\varphi^2)
+\frac{2 U l_p^3 }{N}[d\phi+ { N \over 2 } (\cos\theta -1) d\varphi ]^2
\ee
where $\phi \equiv x_{11}/R_{11}$ has period $\phi \sim \phi + 2 \pi$. 
Defining the new variables $y^2 = 2 N l_p^3 U $, $~ \tilde \theta = \theta/2$,
$~ \tilde \varphi = \varphi + \phi/N$, $~ \tilde \phi = \phi/N$ we 
obtain the metric
\be\label{An}
ds^2=dx^2_{||}+ dy^2 + y^2 (d\tilde \theta^2 +
  \sin^2\tilde \theta d\tilde\varphi^2 +
\cos^2\tilde \theta d\tilde \phi^2 ),
\ee
where $0\leq \tilde \theta \leq \pi/2$ and $0 \leq \tilde \varphi,
\tilde \phi \leq 2\pi$ with the identification $(\tilde \varphi, 
\tilde \phi) \sim (\tilde \varphi, \tilde \phi) + ( 2\pi/N, 2\pi/N)$.
This identification leads  to an 
 ALE space with an $A_{N-1}$ singularity \cite{gh}.
Note that we are saying that an ALE singularity in M-theory 
 has, for large $N$, 
a region which is properly described by a type IIA solution. 

The Riemann curvature tensor  of (\ref{An})  vanishes and the 
identifications involve circles of large proper length if $ y \gg l_p N$
 (i.e. $ \gym^{2/3}  U \gg N$). 
This means that unlike the cases analyzed so far 
 the 11D supergravity solution can be
trusted in the UV {\em for any $N$}.
Since the large $U$ solution is just flat eleven dimensions, 
we do not expect to find  any seven dimensional field theory in the UV which 
flows, in the IR, to 
super-Yang-Mills in $6+1$.
In particular, this implies that DLCQ of M-theory on $T^6$ is as
complicated as M-theory \cite{sen,sei2}.

Another way to state the difference between this case and the previous ones
 is to observe that, in the present case, massive geodesics 
can escape all the way to infinity. There is a second asymptotic
 region which is 
eleven dimensional and is described by M-theory itself. 
For other branes (except for NS 5 branes)
all massive geodesics either fall back into the small $U$ region (IR)
or the supergravity solution is invalid for large $U$ and is replaced by
perturbative super-Yang-Mills (for $p<3$). 
In the Anti-deSitter cases  the only 
geodesics that
reach  infinity are the massless geodesics. In the quantum
 description
one sees that only s-waves can propagate to infinity. 

\subsection{Non-extremal}

In order to analyze this decoupling problem more closely we consider
a near extremal configuration. We  consider a system
of D6 branes at finite temperature  in the limit (\ref{limD6})
and  analyze the corresponding supergravity solution.
For small   energy densities above extremality we are in the
super-Yang-Mills regime. For intermediate energy densities we have
a description involving 
 the type IIA supergravity solution (\ref{nearextr}).
However, if the energy
densities above extremality are large, $\epsilon \gg N l_p^{-7}$, we 
should use the eleven dimensional description. 
As we saw the 11D supergravity can be trusted for any $N$ in the UV.
Starting with the near extremal D6 brane solution (\ref{nearextr})
 and ``uplifting'' it,
as we did for the extremal one, we find a metric which corresponds to 
the metric of an uncharged Schwarzschild black hole sitting at the ALE 
singularity.
More explicitly we get the metric
\be
ds^2=- ( 1 - {y_0^2 \over y^2 }) dt^2 + { dy^2 \over 
( 1 - {y_0^2 \over y^2 }) }
 + y^2 ( d \tilde \theta^2 + 
 \sin^2\tilde \theta d\tilde\varphi^2 +
\cos^2\tilde \theta d\tilde \phi^2 ) + dx_i^2
\ee
where $i=1,..,6$ and the angles have the same identifications that they had
before. The parameter $y_0$ is related to the energy above extremality 
via the  formula $y_0^2 = 2 N l_p^3 U_0$ with  $U_0$  given
in terms of the energy density as in  (\ref{enerdens}).
The Hawking temperature is  $T_H \sim 1/\sqrt{N l_P^9 \epsilon} $. 
We find, therefore,  that we have Hawking radiation into 
the asymptotic region
where we have bulk eleven dimensional supergravity. Thus we conclude 
as in \cite{sei2,sen} 
that there is no decoupled limit for the D6 brane theory.

The main difference between the discussion in this sub-section and
\cite{mal2} is that here the supergravity description 
can be trusted for any $N$ while 
 the supergravity solution found in \cite{mal2} is valid only for large $N$
(and large energy density as in our case).

\bl {\em Conclusions:}

\bl For any $N$ the  UV region is described  by M-theory on a flat
background (an $A_{N-1}$ singularity).
Note that there is  no field theory description in the UV for any $N$.
 This theory flows in the IR to super-Yang-Mills.
For large $N$ there is an intermediate region which can be
described by the IIA D-sixbrane solution. 



{\bf Acknowledgments}

J. M. would like  to thank N. Seiberg, A. Strominger and E. Witten for
nice discussions and Y. Oz and A. Nudelman for pointing out typos in
a previous version of the paper.  
J. M. was supported in part by 
DOE grant
DE-FG02-96ER40559.
The work of N. I., J. S. and S. Y. was  supported
    in part by the US-Israel Binational Science Foundation, by GIF --
    the German-Israeli Foundation for Scientific Research, and by the
    Israel Science Foundation.

\end{document}